\documentclass[doublecol]{epl2}

\newcommand{\ci}{CeIn$_{3-x}$Sn$_x$~}
\newcommand{\cri}{CeRhIn$_{5-x}$Sn$_x$~}

\title{Quantum criticality in layered \cri compared with cubic \ci}
\shorttitle{Title} 

\author{J.G. Donath\inst{1} \and F. Steglich\inst{1} \and E.D. Bauer\inst{2} \and F. Ronning\inst{2} \and J.L. Sarrao\inst{2} \and P. Gegenwart\inst{3}}

\institute{
  \inst{1} Max-Planck-Institute for Chemical Physics of Solids,
D-01187 Dresden, Germany\\
  \inst{2} Los Alamos National Laboratory, Los Alamos, NM 87545,
  USA\\
  \inst{3} I. Physik. Institut, Georg-August-Universit\"{a}t
G\"ottingen, D-37077 G\"ottingen

} 

\abstract{We report low-temperature thermal-expansion measurements
on single crystals of the {\it layered} heavy fermion system \cri
($0.3\leq x \leq 0.6$) and compare it with a previous study on the
related {\it cubic} system \ci [R. K\"{u}chler {\it et al.}, Phys.
Rev. Lett. {\bf 96}, 256403 (2006)]. Both systems display a quantum
critical point as proven by a divergent Gr\"uneisen ratio. Most
remarkably, the three-dimensional itinerant model explains quantum
criticality in {\it both} systems, suggesting that the crystalline
anisotropy in \cri is unimportant. This is ascribed to the effect of
weak disorder in these doped systems.}

\begin{document}

\maketitle

Quantum critical points (QCPs) in intermetallic compounds are of
great scientific interest, as they provide the origin of non-Fermi
liquid (NFL) behavior and novel ground states like unconventional
superconductivity (SC). Heavy fermion (HF) systems, i.e. rare-earth
or actinide-based compounds with competing Kondo- and exchange
interactions are prototype systems for the investigation of QCPs,
and different classes of QCPs have been identified
\cite{Naturephysics}. In one class, the observed properties are in
agreement with the predictions of the spin-density-wave (SDW)
theory, which considers the $f$-electrons as itinerant in the entire
regime close to the QCP. In another class of materials (most
prominent examples include CeCu$_{6-x}$Au$_x$ \cite{Schroeder} and
YbRh$_2$Si$_2$ \cite{Custers,Paschen,Gegenwart}) there are strong
indications for a localization-transition of the $f$-electrons due
to the breakdown of Kondo screening at the QCP. SC has been observed
in some but not all compounds close to QCPs and may even occur near
first-order quantum phase transitions (QPTs) like in CeRh$_2$Si$_2$
\cite{Movshovich, Graf} under pressure, which lack any signatures of
NFL behavior.

There are several indications that magnetic anisotropy may be a
crucial parameter for quantum criticality: (i) quasi-two-dimensional
(2D) magnetic fluctuations have been observed at the QCP in
orthorhombic CeCu$_{5.9}$Au$_{0.1}$ \cite{Stockert} with an
anomalous energy over temperature scaling of the dynamical
susceptibility \cite{Schroeder}, which strongly violates the
predictions of the itinerant SDW theory, (ii) a locally critical QCP
has been predicted for the case of 2D magnetic fluctuations
\cite{Si}, (iii) SC in layered CeTIn$_5$ (T=Co, Ir, Rh) occurs at
ten times higher temperatures compared to the cubic relative
CeIn$_3$ \cite{Mathur,Hegger} and (iv) spin-liquid formation among
the local moments, proposed in the presence of strong geometrical
frustration (which may possibly be enhanced in 2D magnetic systems),
may act as competing mechanism against the Kondo-singlet formation
\cite{Senthil,Burdin}.

In order to systematically investigate the relevance of magnetic
anisotropy on quantum criticality, a comparison of cubic CeIn$_3$
with layered CeTIn$_5$ is most promising. The cubic point symmetry
of Ce atoms in the former must lead to isotropic magnetic
fluctuations. By contrast, in CeTIn$_5$ the alternating series of
CeIn$_3$ and TIn$_2$, stacked along the $c$-axis (for the crystal
structures see Fig.~1), is responsible for a strongly 2D character
of the Fermi surface \cite{dhva} and may also lead to quasi-2D
magnetic fluctuations \cite{Kawasaki}, although the magnetic
correlation length in CeRhIn$_5$ above $T_N$ \cite{Bao} as well as
in superconducting CeCoIn$_5$ \cite{Stock} shows only a moderate
anisotropy.

Hydrostatic pressure experiments have been performed on cubic
CeIn$_3$ (N\'{e}el temperature $T_N$ at ambient pressure about
10\,K) as well as layered CeRhIn$_5$ ($T_N=3.8$\,K). In both cases,
the AF ordering vanishes discontinuously as a function of applied
pressure \cite{Kawasaki2008,Park,Knebel2008}, and the nature of the
$f$-electrons, as determined from de Haas-van Alphen (dHvA)
experiments at low temperatures and high magnetic fields, changes
from localized to itinerant at the critical pressure \cite{Settai}.
An important difference between the two systems is that SC in
CeIn$_3$ occurs only in a very narrow pressure regime and below
0.2\,K, whereas $T_c$ values above 2~K are observed in CeRhIn$_5$
between 2 and 4\,GPa. Although the low-$T$ electrical resistivity of
CeIn$_3$ has shown an anomalous exponent of 1.6 \cite{Mathur},
nuclear quadrupole resonance suggests a Landau Fermi liquid ground
state \cite{Kawasaki2008}, and the cyclotron mass derived from dHvA
experiments is constant near the discontinuous QPT \cite{Settai}.
For CeRhIn$_5$ the cyclotron mass $m^\star(p)$ \cite{Settai} and the
coefficient $A(p)$ of $T^2$ behavior in the electrical resistivity
at 15\,T \cite{Knebel 2008} show diverging behavior, suggesting a
field-induced QCP close to $p_c\approx 2.5$\,GPa.
Previously it has been demonstrated, that Sn-doping in \ci
\cite{Kuechler06} as well as \cri \cite{Bauer06,Donath07} leads
to a continuous suppression of AF order without formation of SC
around the QCP. Therefore, and because the {\it same} control
parameter (Sn-doping) is used to tune the QCP \cite{footnote}, these
two systems seem to be ideally suited to perform the desired
comparative study on the effect of lattice anisotropy on quantum
critical behavior. Figure 1 compares the phase diagrams of the two
systems.

\begin{figure}
  \centering
  \includegraphics[width=\linewidth,keepaspectratio]{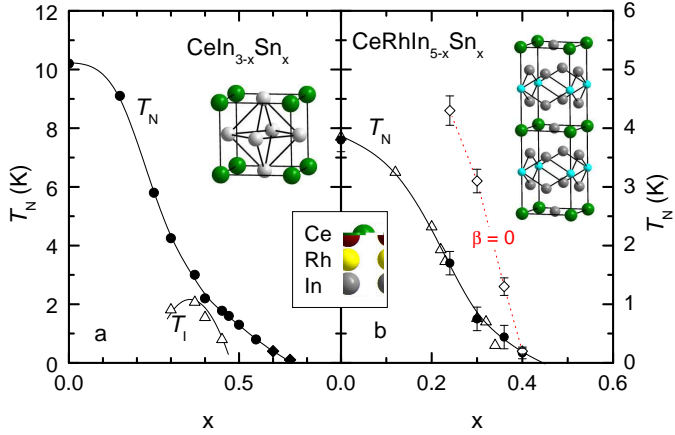}
\caption{(Color online) Magnetic phase diagrams for cubic \ci
\cite{Kuechler06} (a) and tetragonal \cri (b). Open triangles and
closed circles in (b) represent N\'{e}el temperature as determined
from specific-heat \cite{Bauer06} and thermal-expansion
(\cite{Donath07} and this study) measurements. Red dotted line
through open diamonds indicates zero crossing of the volume
expansion with $\beta(T)>0$ ($<0$) on its right (left) side.}
\label{fig1}
\end{figure}

Since Sn has one more $p$-electron compared to In, the partial
substitution of In- by Sn-atoms increases the conduction electron
density of states and thus the Kondo temperature, leading to a
suppression of the AF ordering (a small increase of lattice
constants with Sn-doping is subdominant). For \ci a clear change in
the slope of $T_N(x)$ occurs close to a presumed tetracritical point
at $x\approx 0.4$ \cite{Pedrazzini} beyond which a quasi-linear
suppression of the N\'{e}el temperature towards a QCP at $x_c=0.65$
has been found \cite{Kuechler06}. The phase diagram of \cri as
displayed in Fig.~1b is rather similar and also shows a change of
the $T_N(x)$ slope before the QCP is reached. Previous
low-temperature specific heat and electrical resistivity
measurements down to 0.4\,K suggest a QCP near $x=0.4$. Our
thermal-expansion measurements at $T\geq\,0.08$\,K, discussed below,
reveal $x_c\simeq 0.46$.


For this study, we have used the same single crystals studied in
\cite{Bauer06,Donath07}, as well as $x=0.44$ and $x=0.6$ single
crystals prepared similarly. We always refer to the actual Sn
concentration $x$ determined by microprobe analysis with an
uncertainty of less than $1\,\%$. The residual resistivity of the
\cri crystals increases monotonically with $x$ and reaches
$28\,\mu\Omega$cm at $x=0.48$ \cite{Bauer06}. Extended X-ray
absorption fine structure measurements in CeCoIn$_{5-x}$Sn$_x$ have
revealed that the Sn atoms preferentially occupy the In-(1) position
within the CeIn$_3$ planes of the layered system \cite{Daniel}.
Similar behavior arises in \cri \cite{Rusz}. The linear thermal
expansion $\alpha(T)=d[\Delta L(T)/L]/dT$ has been determined with
the aid of a high-resolution capacitive dilatometer, attached to a
dilution refrigerator. The volume-expansion coefficient displayed in
Figure 2 has been determined by $\beta=\alpha_{\parallel
  c}+2\alpha_{\perp c}$.


\begin{figure}
  \centering
  \includegraphics[width=\linewidth,keepaspectratio]{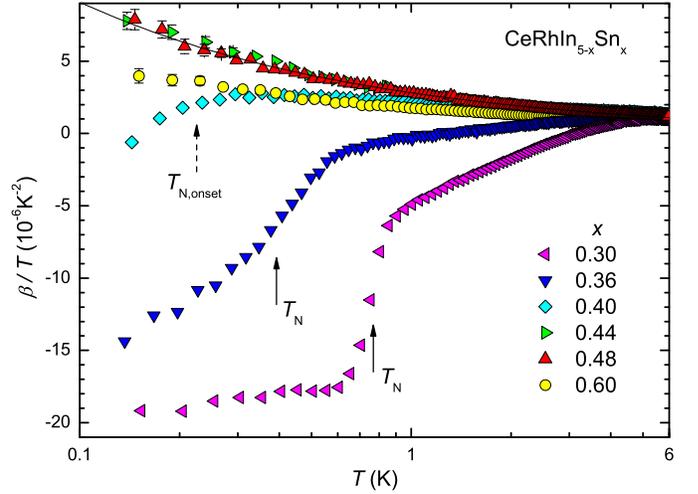}
  \caption{(Color online) Temperature dependence of the volume thermal
expansion coefficient $\beta=\alpha_{\parallel c}+2\alpha_{\perp c}$
of CeRhIn$_{5-x}$Sn$_x$ as $\beta/T$ vs $T$ (on a logarithmic
scale). Solid and dotted arrows indicate $T_N$ and onset of
(broadened) N\'{e}el transition, respectively. Line displays
$\beta(T)/T=b_0+b_1/\sqrt{T}$ dependence with $b_0=-0.08\times
10^{-6}$K$^{-2}$ and $b_1=2.9\times 10^{-6}$K$^{-1.5}$.}
\label{fig2}
\end{figure}

Thermal-expansion measurements on CeRhIn$_{5-x}$Sn$_x$ to
investigate the long-range antiferromagnetism have been discussed
previously \cite{Donath07}. For $x\leq 0.24$, a positive
discontinuity $\Delta\beta>0$ has been observed at the N\'{e}el
temperature, reflecting an increase of $T_N$ with hydrostatic
pressure. These samples are thus located on the left side of the
maximum of $T_N(P)$ expected within the Doniach diagram. Beyond
$x=0.24$, where the change in slope in $T_N(x)$ occurs (cf.
Fig.~1b), $\Delta\beta<0$, indicating that the system approaches the
QCP. A change of sign in the volume thermal expansion $\beta(T)$
which indicates an accumulation point of entropy \cite{Garst} occurs
for \ci very close to $T_N(x)$ \cite{Kuechler06}. By contrast it is
located much above the N\'{e}el temperature for layered \cri (cf.
the red dotted line in Fig.~1b). This may indicate a largely
extended Ginzburg regime in which classical critical fluctuations
dominate. However, specific-heat measurements have shown a
Schottky-like anomaly in $C(T)/T$ very close to this line. Thus, an
additional energy scale exists in this system which is likely
related to short-range magnetic correlations \cite{Bauer06}. Most
interestingly, it also vanishes in the vicinity of the QCP, i.e. in
the range $0.4<x<0.44$ (cf. Fig.~2). The QCP is located in between
the concentrations $x=0.44,$ and $x=0.48$ for which, within
experimental resolution, the same divergent behavior in $\beta(T)/T$
is found down to the lowest temperatures. At higher Sn content,
$x=0.60$, $\beta(T)/T$ tends to saturate at lowest temperatures,
indicative for a crossover towards Landau Fermi liquid behavior.
In the following, we will analyze quantum criticality for $x=0.48$.

\begin{figure}
  \centering
  \includegraphics[width=\linewidth,keepaspectratio]{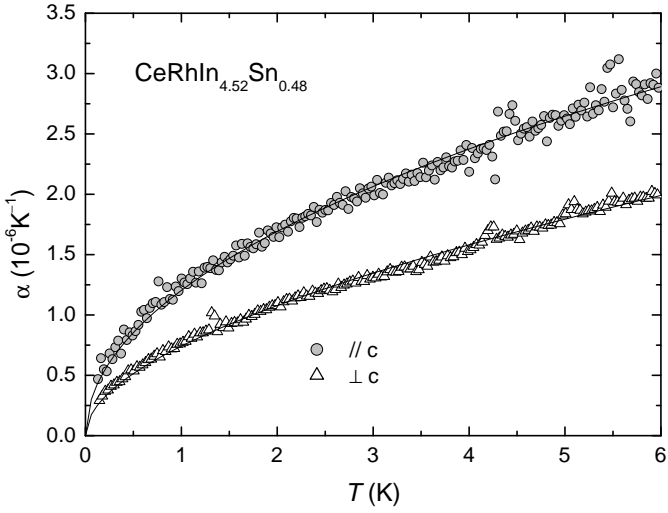}
\caption{Linear thermal expansion coefficient of
CeRhIn$_{4.52}$Sn$_{0.48}$ along (circles) and perpendicular
(triangles) to the $c$-axis. Lines indicate
$\alpha(T)=a_0T+a_1\sqrt{T}$ with $a_0=-0.01\times 10^{-6}$K$^{-2}$
($0.05\times 10^{-6}$K$^{-2}$) and $a_1=1.2\times 10^{-6}$K$^{-1.5}$
($0.7\times 10^{-6}$K$^{-1.5}$) for $\alpha\parallel c$ ($\alpha
\perp c$). The noise for $\alpha_\parallel$ is relatively large,
because the sample length along the $c$-axis is rather small.}
\label{fig3}
\end{figure}

\begin{figure}
  \centering
  \includegraphics[width=\linewidth,keepaspectratio]{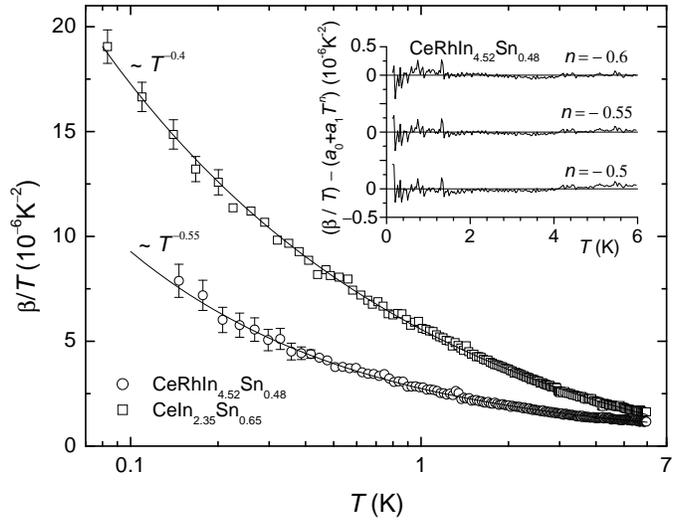}
\caption{Volume expansion coefficient as $\beta/T$ vs $T$ on
logarithmic scale for CeRhIn$_{4.52}$Sn$_{0.48}$ (circles) and
CeIn$_{2.35}$Sn$_{0.65}$ (squares, \cite{Kuechler06}). Solid lines
indicate power-law behavior. The inset displays the deviation of
$\beta(T)/T$ data for CeRhIn$_{4.52}$Sn$_{0.48}$ from the best-fit
description $\beta(T)/T=a_0+a_1T^{n}$ using different exponents $n$.
The constant term is negligible in all cases, $|a_0|\leq 0.2\times
10^{-6}$K$^{-2}$. For clarity, the three data sets have been shifted
by different amounts vertically.} \label{fig4}
\end{figure}

Figure 3 displays measurements of the linear thermal-expansion
coefficient of CeRhIn$_{4.52}$Sn$_{0.48}$ along and perpendicular to
the $c$-axis. Despite moderate anisotropy
($\alpha_\parallel/\alpha_\perp=1.7$), similar behavior is found
along both directions, namely a square-root behavior in $\alpha(T)$.
Such a temperature dependence is expected within the itinerant SDW
theory for a 3D AF QCP \cite{Zhu}. By contrast,
$\alpha(T)\approx const$ is expected in the 2D case. 
Further evidence for the 3D nature of quantum criticality in \cri is
provided by the analysis of the volume thermal expansion and
Gr\"uneisen ratio and comparison with the case of cubic \ci.
In the latter case the AF QCP is located at $x=0.65$, and the volume
thermal expansion has been described by $\beta(T)/T=a_0+a_1T^n$,
yielding $n=-0.4$ for a fit in the temperature range 0.1\,K$\leq T
\leq$\,6\,K and $n=-0.5$ for a fit at temperatures below 1\,K
\cite{Kuechler06}. A value of $n=-0.5$ agrees with the prediction of
the SDW theory for a 3D AF QCP, whereas $n=-1$ is expected for the
2D case \cite{Zhu}. In Figure~4, we compare $\beta/T$ for
CeIn$_{2.35}$Sn$_{0.65}$ with respective data on
CeRhIn$_{4.52}$Sn$_{0.48}$. The best-fit description of the latter
system reveals an exponent $n=-0.55\pm 0.05$ (cf. the inset which
displays the deviation from power law fits with variable exponent).
Thus, thermal expansion does not reveal a significant difference
between quantum criticality in layered \cri compared to cubic
CeIn$_{3-x}$Sn$_x$.

\begin{figure}
  \centering
  \includegraphics[width=\linewidth,keepaspectratio]{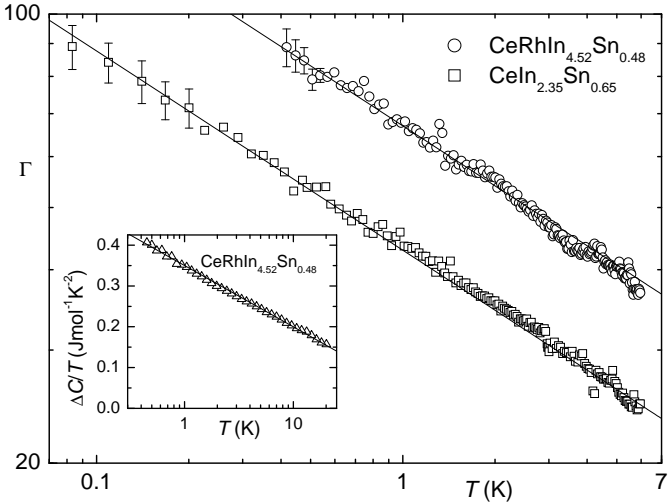}
\caption{Temperature dependence (on double-log scales) of the
Gr\"uneisen parameter $\Gamma=(V_{mol}\beta)/(\kappa_T\Delta C)$ of
CeRhIn$_{4.52}$Sn$_{0.48}$ and CeIn$_{2.35}$Sn$_{0.65}$
\cite{Kuechler06}, where $V_m$ and $\kappa_T$ denote the molar
volume and the isothermal compressibility, respectively. Solid lines
display $T^{-0.31}$ behavior. The inset displays the specific heat
increment $\Delta C=C-C_{Phonon}$ of CeRhIn$_{4.52}$Sn$_{0.48}$ as
$\Delta C/T$ vs $T$ on a logarithmic scale. The phonon contribution
has been determined from data on LaRhIn$_5$ \cite{Bauer06}.}
\label{fig5}
\end{figure}


The specific heat of \cri has been investigated down to 0.4\,K
\cite{Bauer06}. As shown in the inset of Figure 5, $C(T)/T$ follows
a $\log T$ dependence between 0.4 and 20\,K, similar as observed in
many other NFL systems \cite{Stewart1,Stewart2}. This temperature
dependence is expected within the 3D SDW theory in an intermediate
range \cite{Moriya}, while at lowest temperatures a crossover to a
square-root temperature dependence is predicted. Such an expected
crossover has indeed been found in CeIn$_{2.35}$Sn$_{0.65}$ around
0.4\,K \cite{Radu}, although below 0.2\,K the specific heat is
dominated by the nuclear quadrupolar contribution of indium. As
already discussed for the case of CeCoIn$_{5-x}$Sn$_x$ \cite{Donath
08}, thermal expansion and the Gr\"uneisen parameter are thus
ideally suited to investigate the nature of quantum criticality in
these systems as they mitigate the obscuring effects of the nuclear
contribution to specific heat. Theory predicts a stronger than
logarithmic divergence of the Gr\"uneisen paramater $\Gamma(T)$ for
any pressure-sensitive QCP \cite{Zhu}, otherwise, as recently found
in CePd$_{1-x}$Rh$_x$~\cite{Westerkamp}, quantum criticality as
source of NFL behavior could be excluded. As shown in Figure~5,
which displays $\Gamma(T)$ on double-logarithmic scales, such
stronger-than $\log T$ divergence is indeed present for
CeRhIn$_{4.52}$Sn$_{0.48}$ and CeIn$_{2.35}$Sn$_{0.65}$. Most
interestingly, a very similar $T$-dependence is found for the two
systems. In order to compare with the theoretical predictions, the
{\it critical} Gr\"uneisen ratio
$\Gamma^{cr}\propto\beta^{cr}/C^{cr}$ must be analyzed, where
$\beta^{cr}$ and $C^{cr}$ denote the volume thermal expansion and
specific heat after subtraction of non-critical, i.e., Fermi-liquid
like contributions. Within the 3D SDW model, the noncritical
contribution to specific heat is given by the saturation value of
$C(T)/T$ as $T\rightarrow 0$ \cite{Zhu}. Assuming a saturation of
the specific heat coefficient at either 0.5, 0.55 or
0.6\,Jmol$^{-1}$K$^{-2}$, respectively would yield values of $-1.2$,
$-1$ or $-0.93$ for the Gr\"uneisen exponents within the temperature
interval 0.4\,K$\leq T\leq$\,6\,K. For CeIn$_{2.35}$Sn$_{0.65}$, a
value of $-1.1\pm 0.1$ has been found \cite{Kuechler06}. Thus, the
critical Gr\"uneisen analysis also suggests strong similarities in
quantum critical behavior of the two systems.

Weak disorder may strongly influence the nature of quantum
criticality and the dimensionality of the critical fluctuations,
whereas strongly disordered systems like UCu$_{5-x}$Pd$_x$ do not
display a QCP, and NFL behavior in such systems appears to be
disorder-driven \cite{Miranda}. Previously, we have investigated the
influence of disorder on quantum criticality in CeCoIn$_{5-x}$Sn$_x$
\cite{Donath08}. With increasing $x$, a field-tuned QCP remains
pinned to the upper critical field $H_{c2}(x)$ of heavy-fermion
superconductivity, which is linearly suppressed to 0 at $x=0.18$
\cite{Bauer}. While the specific heat remains virtually unchanged
with $x$ at the respective critical fields, thermal expansion
delineates a crossover scale $T^\star(x)$ separating 2D from 3D
quantum critical behavior \cite{Donath08}. This crossover scale
increases from 0.3\,K at $x=0$ to 1.4\,K at $x=0.18$ with increasing
disorder ($x$), characterized by a residual resistivity $\rho_0 = 15
\mu\Omega$cm for $x=0.18$.
%
In \cri, even four-times higher Sn-concentrations (resulting in
$\rho_0=28\,\mu\Omega$cm) are required to access the QCP. We
conjecture that 3D behavior therefore extends up to at least 6\,K.
This interpretation assumes that isotropic impurity scattering due
to the In-Sn site disorder is effective in smearing out the
anisotropy of the quantum critical fluctuations. Nevertheless, as
evidenced by the divergent Gr\"uneisen ratio, a truely
pressure-sensitive QCP emerges. This is in contrast to strongly
disordered systems mentioned above which do not display a QCP.
Furthermore, we note that $\rho_0\approx 40\,\mu\Omega$cm of
CeCu$_{5.9}$Au$_{0.1}$ \cite{Lohneysen} is even $40\%$ larger as in
CeRhIn$_{4.52}$Sn$_{0.48}$ but CeCu$_{5.9}$Au$_{0.1}$ nevertheless
displays quasi-2D quantum critical fluctuations \cite{Stockert}.
This indicates that the CeMIn$_5$ systems are rather sensitive to
disorder within the tetragonal CeIn$_3$ plane and, in general, that
weak disorder can drastically influence quantum criticality.

%
%


We conclude by stating that weak disorder as introduced by low
Sn-doping in layered CeRhIn$_5$ stabilizes {\it threedimensional}
quantum critical behavior. Thus, no significant differences in the
low-$T$ thermal expansion and Gr\"uneisen ratio to the corresponding
quantities in cubic \ci could be resolved. The distinct effect of
weak disorder on the quantum criticality in the 115 systems appears
to be independent of the experimental tuning parameter, as this had
already been observed at the {\it magnetic field-induced QCP} in
Sn-doped CeCoIn$_5$ \cite{Donath08}.

\acknowledgments Work at Dresden and G\"{o}ttingen was partially
financed by the DFG Research unit 960 (Quantum phase transitions),
while work at Los Alamos was carried out under the auspices of the
U.S. DOE.

\end{document}